\documentclass[acmtog, nonacm]{acmart}

\AtBeginDocument{%
  \providecommand\BibTeX{{%
    \normalfont B\kern-0.5em{\scshape i\kern-0.25em b}\kern-0.8em\TeX}}}

\usepackage{amsmath,amsfonts}
\usepackage{algorithmic}
\usepackage{graphicx}
\usepackage{textcomp}
\usepackage{amsthm}
\usepackage{amsmath}
\usepackage{caption}
\usepackage{subcaption}
\usepackage[T1]{fontenc}
\usepackage[utf8]{inputenc}
\usepackage{epstopdf}
\usepackage{enumerate}
\usepackage{lipsum}
\usepackage{gensymb}
\usepackage{etoolbox}
\usepackage{graphics}
\begin{document}

\title{Using Multivariate Linear Regression for Biochemical Oxygen Demand Prediction in Waste Water}

\author{Isaiah K. Mutai}
\email{mutaikiprono1@gmail.com}
\authornotemark[1]
\affiliation{%
  \institution{Department of Chemical Engineering, Dedan Kimathi University of Technology}
  \streetaddress{P.O. Box 657 }
  \city{Nyeri}
  \country{Kenya}
  \postcode{10100}}

\author{Kristof Van Laerhoven}
\affiliation{%
  \institution{Department of Ubiquitous Computing, University of Siegen}
  \streetaddress{Holderlin str.57076}
  \city{Siegen}
  \country{Germany}}

\author{Nancy W. Karuri}
\affiliation{%
  \institution{Department of Chemical Engineering, Dedan Kimathi University of Technology}
  \city{Nyeri}
  \country{Kenya}}

\author{Robert K. Tewo}
\affiliation{%
 \institution{Department of Chemical Engineering, Dedan Kimathi University of Technology}
 \streetaddress{P.O. Box 657}
 \city{Nyeri}
 \country{Kenya}}
 



\renewcommand{\shortauthors}{Mutai, et al.}

\begin{abstract}
  There exist opportunities for Multivariate Linear Regression (MLR) in the prediction of Biochemical Oxygen Demand (BOD) in waste water, using the  diverse water quality parameters as the input variables. The goal of this work is to examine the capability of MLR in prediction of BOD in waste water through four input variables: Dissolved Oxygen (DO), Nitrogen, Fecal Coliform and Total Coliform. The four input variables have higher correlation strength to BOD out of the seven parameters examined for the strength of correlation. Machine Learning (ML) was done with both 80\% and 90\% of the data as the training set and 20\% and 10\% as the test set respectively. MLR performance was evaluated through the coefficient of correlation (r), Root Mean Square Error (RMSE) and the percentage accuracy in prediction of BOD. The performance indices for the input variables of Dissolved Oxygen, Nitrogen, Fecal Coliform and Total Coliform in prediction of BOD are: RMSE=6.77mg/L, r=0.60 and accuracy 70.3\% for training dataset of 80\% and RMSE=6.74mg/L, r=0.60 and accuracy of 87.5\% for training set of 90\% of the dataset. It was found  that increasing the percentage of the training set above 80\% of the dataset improved  the accuracy of the model only but did not have a significant impact on the prediction capacity of the model. The results showed that MLR model could be successfully employed in the estimation of BOD in waste water using appropriately selected input parameters.
\end{abstract}



\maketitle

\section{Introduction}
{W}{ater} is the solvent necessary for life  to be in existence and to develop. It is among others, the first item that makes the earth to be habitable and hence unique from the other planets in the solar system\cite{Lammer2009}. Water however, is a scarce resource with less than 1\% of the water on the surface of the earth being usable and available as freshwater\cite{WMO2021}. It is projected that over 5 billion people will suffer water scarcity by the year 2050\cite{UN-Water2018}. This scarcity can be attributed to the increase in the demand for water estimated at the rate of 1.8\% per year and also an increasing world population  expected to reach 9.4-10.2 billion people by the year 2050\cite{Boretti2019}. Recycling of waste water for reuse is important in reducing this problem of water scarcity\cite{Tzanakakis2020}. 

Machine Learning is among today’s fastest growing fields. It is projected to emerge as the most transformative technology of the 21st century, hence, the need for its utilization\cite{Jordan2015}. Although it has registered success in a number of sectors, ML remains underexploited in the field of waste water treatment \cite{Jordan2015,Wang2021}. Large volumes of datasets are generated in Waste Water Treatment Plants (WWTP), but the utilization of these data is low owing to the lack of background in data science among water treatment professionals \cite{Newhart2019}.

 The main contaminant in waste water is the organic matter  \cite{Jouanneau2019}. Monitoring the amount of the organic matter in waste water is therefore key and paramount to ensure the appropriate treatment measures are put in place. This depends on the extend of pollution of the water. BOD measurement offers the option to achieve this objective and is used as one of the water quality indices \cite{Yu2019}. BOD$_5$ which is the standard BOD measurement is time consuming. It takes 5 days to give the results of the measurement, hence there is a danger of delayed mitigation action against pollution \cite{Ooi2022}. Machine Learning can bridge this gap by predicting BOD$_5$ based on input of a  few water quality parameters within a few hours. Multivariate Linear Regression (MLR) is considered a conventional method for water quality parameter prediction \cite{Bilali2020}. In this study, MLR was applied to offer BOD prediction using four input parameters namely: Dissolved Oxygen (DO), Nitrogen, Fecal Coliform and Total Coliform. 

The contribution of this work is twofold: (1) It was found out that  among the key water quality parameters, there is  a strong correlation between Dissolved Oxygen, Fecal Coliforms, Total Coliforms and Nitrogen to BOD and (2) the work demonstrated that better performance capacity of MLR is pegged on the choice of the input parameters.

\section{Materials and Methods}
\begin{table*}[ht]
    \caption{The structure of the dataset \protect\cite{Agrawal2020} used in this paper consists of 64 bit floating point values that sometimes are missing, hence we list below the number of valid measurements: (a)before imputation and (b) after imputation.}
    \begin{subtable}{.5\linewidth}
      \centering
        \caption{Structure of data before imputation}
        \begin{tabular}{ll}
        \begin{tabular}{|l|c|l|l|c|l|}
\hline
          Column    & Valid Measurements &  Dtype & \% missing \\
          \hline
               TEMP &   529  &  float64 &    0.94\\
             
                 DO &   532  &  float64 &    0.37\\
                
                 pH &   534  &  float64 &    0.00\\
                 
       CONDUCTIVITY &   504  &  float64 &    5.62 \\
       
                BOD &   528  &  float64 &    1.12\\
                
NITRATE\_N\_NITRITE\_N &   532  &  float64 & 0.37 \\

     FECAL\_COLIFORM &   452  &  float64  &  15.36\\
     
     TOTAL\_COLIFORM &   495  &  float64 &   7.30\\
     \hline
\end{tabular}
        \end{tabular}
    \end{subtable}%
    \begin{subtable}{.5\linewidth}
      \centering
        \caption{Structure of data after imputation}
       \begin{tabular}{|l|c|l|l|}
\hline
          Column    & Valid Measurements  &  Dtype  \\
          \hline
               TEMP &   534  &  float64 \\
                 DO &   534  &  float64 \\
                 pH &   534  &  float64 \\
       CONDUCTIVITY &   534  &  float64 \\
                BOD &   534  &  float64 \\
NITRATE\_N\_NITRITE\_N & 534  &  float64 \\
     FECAL\_COLIFORM &   534  &  float64 \\
     TOTAL\_COLIFORM &   534  &  float64 \\
     \hline
\end{tabular}
    \end{subtable} \\
\end{table*}
\subsection{Experimental Data}
The data used in this work constitutes a dataset for water quality parameters for different rivers in India. The data provides eight water quality parameters. The values of each parameter is the average taken over a period of time as compiled from the official website for data related to India\cite{Agrawal2020}. The water quality parameters provided in the dataset are: Temperature, Dissolved Oxygen, pH, Conductivity, BOD, Nitrogen, Fecal Coliform and Total Coliform. The few missing data values in the dataset was addressed through data preprocessing, by imputing with the mean values of the column features. Nitrogen is in the form of both Nitrate and Nitrite (Nitrate$\_$N and Nitrite$\_$N).

\subsection{Data Preprocessing and Visualization}
Data Preprocessing was done by dropping off  all non-numeric data and accounting for the missing data values by imputing them with column feature mean values. After the preprocessing, the  visualization of the data was done by means of Principal Component Analysis (PCA) and t-Distributed Stochastic Neighbour Embedding (t-SNE) analysis.
\subsection{Parameter Selection}
The selection of the parameters used in the Machine Learning was done through the analysis of the strength of association of each parameter to the target variable (BOD). This was achieved through the analysis of Pearson correlation of each independent variable to BOD in addition to the findings obtained from PCA analysis. The parameters that showed strong correlation were chosen for the machine learning while the parameters with weaker correlation with BOD were discarded. 

\subsection{Machine Learning}
Machine Learning was done by use of  Multivariate Linear Regression (MLR) using the Linear model imported from the scikit learn library (Sklearn). The training was done with both 80\%/20\% and 90\%/10\%  training/test data of the whole dataset.

Equation (1) shows the the general form of the  MLR model, where: y is the dependent variable, $\beta_{0}$ is the y-intercept, $C_1,C_2,C_3....C_n$ are the coefficients for the independent variables and $x_1,x_2,x_3...x_n$ are the independent variables of the model.
\begin{equation}
    \mathrm{y} =  \mathrm{C}_{1}  \mathrm{x}_{1}
        + \mathrm{C}_{2}  \mathrm{x}_{2}
        + \mathrm{C}_{3}  \mathrm{x}_{3} 
        + \mathrm{C}_{4}  \mathrm{x}_{4}
           \mathrm{...}
        +\mathrm{C}_{n}   \mathrm{x}_{n}
        +\beta_{0}        \label{eqn1}
    \end{equation}

\subsection{Evaluation Criteria}
The evaluation criteria adopted in this study for evaluation of the model performance was the coefficient of correlation (r), the Root Mean Squared Error (RMSE) and accuracy. The coefficient of correlation (r) is a key common criterion for checking the goodness of the line of best fit\cite{Abyaneh2014}. It checks the fitness of the regression model to the data rather than the capability of the model in prediction. A well-fitting model results in predictions being close to the observed values. Nonetheless, it is worth noting that the coefficient of correlation does not work well for all data and hence cannot be relied on as the only measure of performance of the prediction model\cite{Razi2005}. RMSE on the other hand indicates the absolute fitness of the model to the data and has the advantage of being expressed in the same units as the response variable. It is the best criterion for a fit when the main reason for the model is prediction. When RMSE is low and r is high, the model is considered to be good\cite{Guclu2010,Martin2019}.

Accuracy also gives a glimpse of the performance of a model. It is however not a dependable option for the evaluation of the model performance\cite{Vallantin2018}. Algorithms with lower accuracy could be preferred over those with higher accuracy upon consideration of the other factors of performance\cite{Webb2001}. The average  prediction accuracy  for a model to be acceptable is 50\% \cite{Lerios2019}. An accuracy of 70\%-90\%  is an  excellent range which is  consistent with the commonly used industrial standard\cite{Barkved2022}.

\section{RESULTS AND DISCUSSION}

The summary of the data  obtained in its original form is shown in Table 1(a). As can be seen in Table 1(a), it is clear that some measurements are missing. The size of the missing data values is however negligible, with only one case exceeding 10\%. This  marked the highest value of missing measurements (15\%), which is equivalent to 82 measurements. Table 1(b), shows the summary after imputation of the missing data with the mean values of each parameter measured.

\begin{figure}[t]
  \includegraphics[width=\columnwidth]{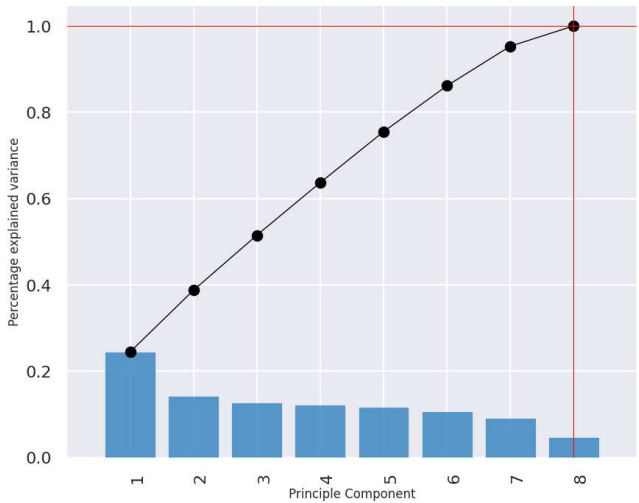}
  \caption{A scree plot showing the explained variance with respect to the principal components. All the principal components contributed almost equally in the total accounting of the variation in the the data}
\end{figure}

\begin{figure}[t]
  \includegraphics[width=\columnwidth]{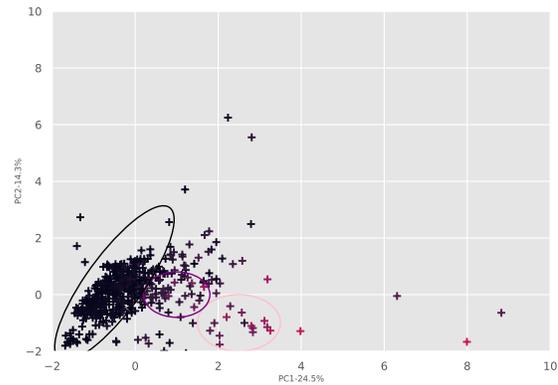}
  \caption{Principal Component Analysis (PCA). The dataset falls into three major clusters with a few data points dispersed outside the clusters in the lower dimension space}
  \label{fig:pca_plot}
\end{figure}

\begin{table*}[t]
\centering
\caption{Pearson Correlation. The summary of all the correlations among all the variables in the dataset is provided }
\resizebox{\textwidth}{!}{%
\begin{tabular}{|l|r|r|r|c|r|c|c|c|}
\toprule
\hline
{} &      TEMP &        pH &        DO &  CONDUCTIVITY &       BOD &  NITRATE\_N\_NITRITE\_N &  FECAL\_COLIFORM &  TOTAL\_COLIFORM \\
\hline
\midrule
TEMP                &  1.000 &        \\
\hline
pH                  &  0.018 &  1.000 \\
\hline
DO                  & -0.185 &  0.066 &  1.000 \\
\hline
CONDUCTIVITY        &  0.074 &  0.012 & -0.105 &      1.000  \\
\hline
BOD                 & -0.071 & -0.056 & -0.522 &      0.099 &  1.000 \\
\hline
NITRATE\_N\_NITRITE\_N &  0.089 & -0.019 & -0.269 &      0.084 &  0.285 &             1.000  \\
\hline
FECAL\_COLIFORM      &  0.004 & -0.013 & -0.080 &     -0.001 &  0.299 &             0.018 &        1.000 \\
\hline
TOTAL\_COLIFORM      & -0.003 & -0.030 & -0.230 &      0.001 &  0.174 &             0.131 &        0.036 &        1.000 \\
\hline
\end{tabular}}
\end{table*}

\begin{table*}[t]
\centering
\caption{The correlation between the individual parameters and the  target variable (BOD). DO, Nitrogen(Nitrate\_N\_Nitrite\_N), Fecal Coliforms and Total Coliforms correlate strongly to BOD while Temperature, pH and Conductivity bear a weak correlation with BOD}
\begin{tabular}{|l|l|l|l|l|l|l|l|}
\hline
{} &            Temp &              pH &             DO &     Conductivity & Nitrate\_N\_Nitrite &      F.Coliforms &     T.Coliforms \\
\hline
Parameters    &           -0.071 &           -0.056 &          -0.522 &              0.099 &              0.285 &              0.299 &            0.174 \\
\hline
Corr\_Strength &  Small Negative &  Small Negative &  High Negative &  Small Posivtive &   Medium Positive &  Medium Positive &  Small Positive \\
\hline
\end{tabular}
\end{table*}

Figure 1, Figure 2 and Figure 3 are the results of data visualization by PCA method. Figure 1 is a scree plot showing the  explained variance  with respect to the principal components. As it can be seen in Figure 1, the bar heights are almost equal, with the exception of the first principal component and the last principal component. This is indicative of the fact that each principal component  contributed almost equally in the total accounting of the variation in the data. The first principal component however, had higher contribution as expected while the last principal component similarly had the least contribution. A PCA plot for the first two principal components is shown in Figure 2. Figure 2 indicated that there were mainly three major clusters of data points when the data was mapped to a lower dimension as indicated by the circled regions of the graph. However, there was an overlap between the clusters as shown. This was an indication of the fact that the parameters were closely associated. The clusters seemed to be oriented towards a direction tilted to the right side of the plot. Figure 3 shows the biplot of the PCA analysis. According to Figure 3, the clusters initially identified in Figure \ref{fig:pca_plot} are oriented in the direction of increasing BOD (low to high) with a loading of 0.569927 on the first principal component. Figure 3 further hinted the existence of a strong and close positive association between BOD, Total coliforms, Nitrogen (Nitrate\_N\_Nitrite\_N) and Fecal Coliforms. The correlation of BOD towards Conductivity and Temperature is a weak positive (angle almost 90\degree) while the association between BOD and pH is a weak negative (angle slightly greater than 90\degree). Dissolved Oxygen (DO) shows a very strong negative correlation to BOD. It is oriented almost in the opposite direction. On the other hand, there is zero linear correlation between Fecal Coliforms and pH (angle = 90\degree)!


\begin{figure}[t]
  \includegraphics[width=\columnwidth]{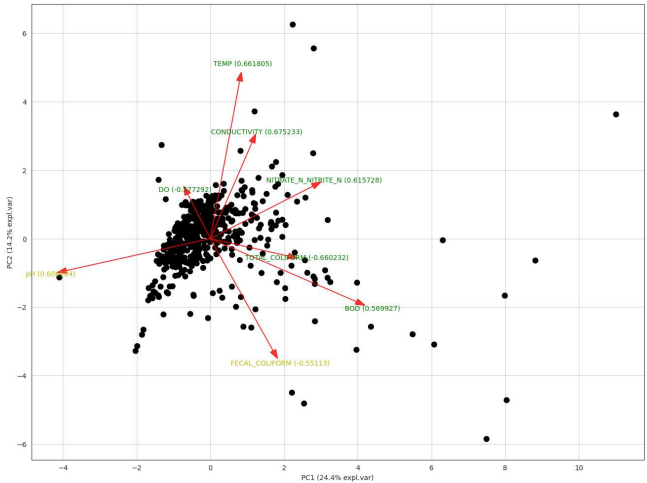}
  \caption{Biplot of the PCA analysis on the first and second principal components. There is  a strong correlation between Dissolved Oxygen, Fecal Coliforms, Total Coliforms and Nitrogen to BOD. pH and Fecal coliforms have no linear correlation  toward each other}
\end{figure}

K-Means clustering as shown in Figure 4 confirmed the existence of three unique clusters as per the elbow analysis technique. This was in agreement with the PCA findings in Figure 2. The t-SNE visualization further shed light on the existing clusters.  Figure 5(a) shows the t-SNE analysis on the target variable of the study (BOD). Figure 5(a) alluded that  the clusters were easily  on the basis of the levels of BOD. This agrees with the PCA findings. It is clear that a larger number of data points fell in the lower level of BOD (0-0.2), then another larger set within the medium level (0.4-0.6) and a small number of data points are within  the high BOD level (0.8-1.0). This is also in agreement with the findings of Figure 3, which indicated that the cluster sizes reduced in size in the direction of increasing BOD. Further t-SNE analysis, Figure 5(b), developed on the basis of the clusters used clearly showed the clusters as low, medium, and high BOD with an indication of an overlap at the transition boundaries. This can be attributed to the possibility of existence of a trend, which could easily fit to a regression model; from low, through the medium and to the high BOD level.

\begin{figure}[t]
  \includegraphics[width=\columnwidth]{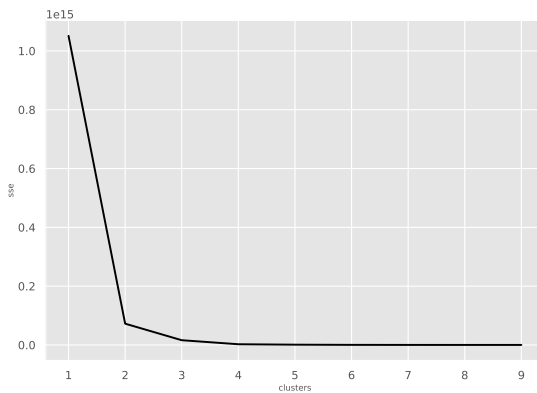}
  \caption{K-Means clustering of the data. The elbow rule shows the existence of 3 unique clusters in the data structure}
\end{figure}

\begin{figure*}[htb!]
\centering
\begin{subfigure}{0.5\textwidth}
  \centering
  \includegraphics[width=\textwidth]{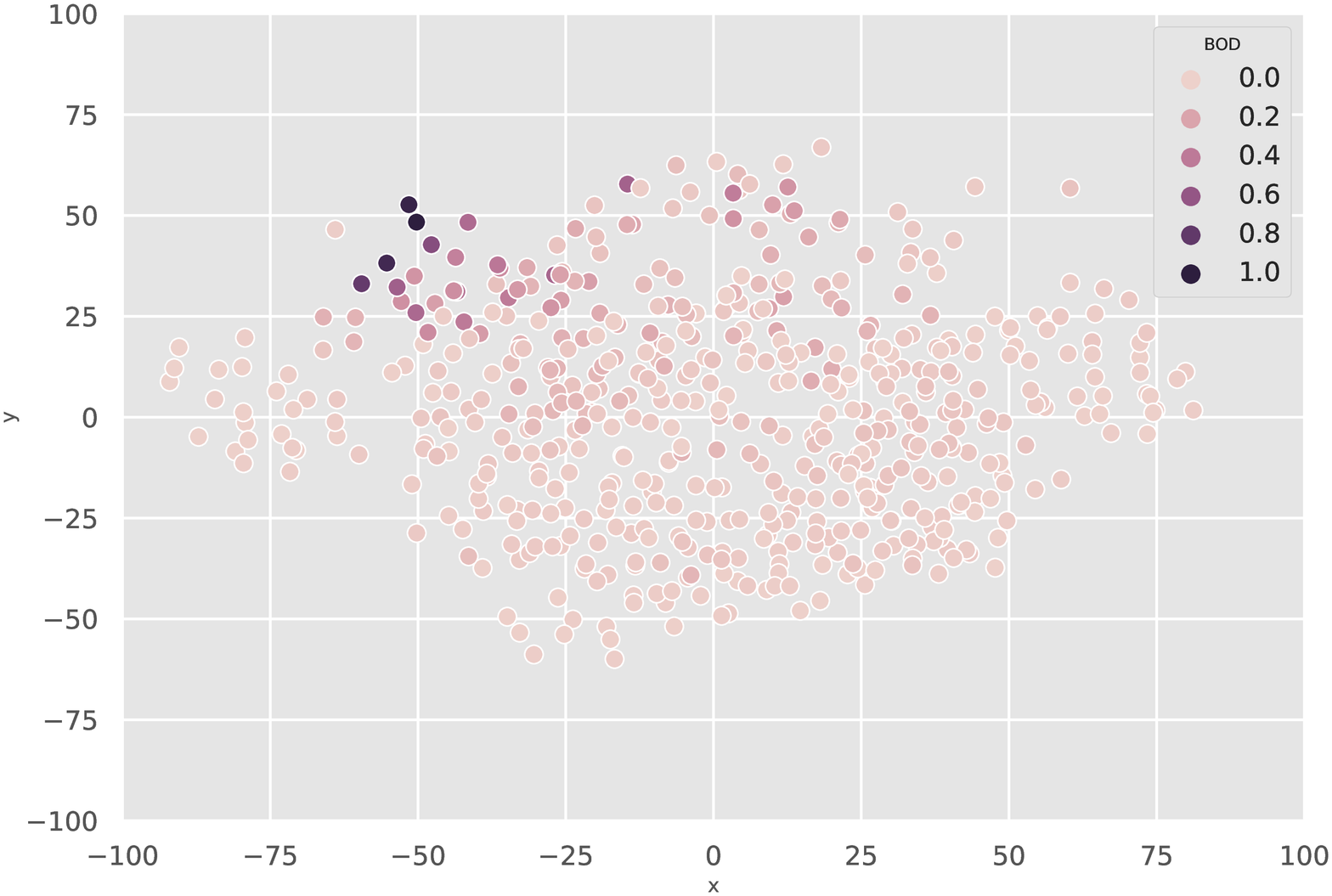}
  \caption{t-SNE with the target being BOD values}
  \label{fig:sub7}
\end{subfigure}%
\begin{subfigure}{0.5\textwidth}
  \centering
  \includegraphics[width=\textwidth]{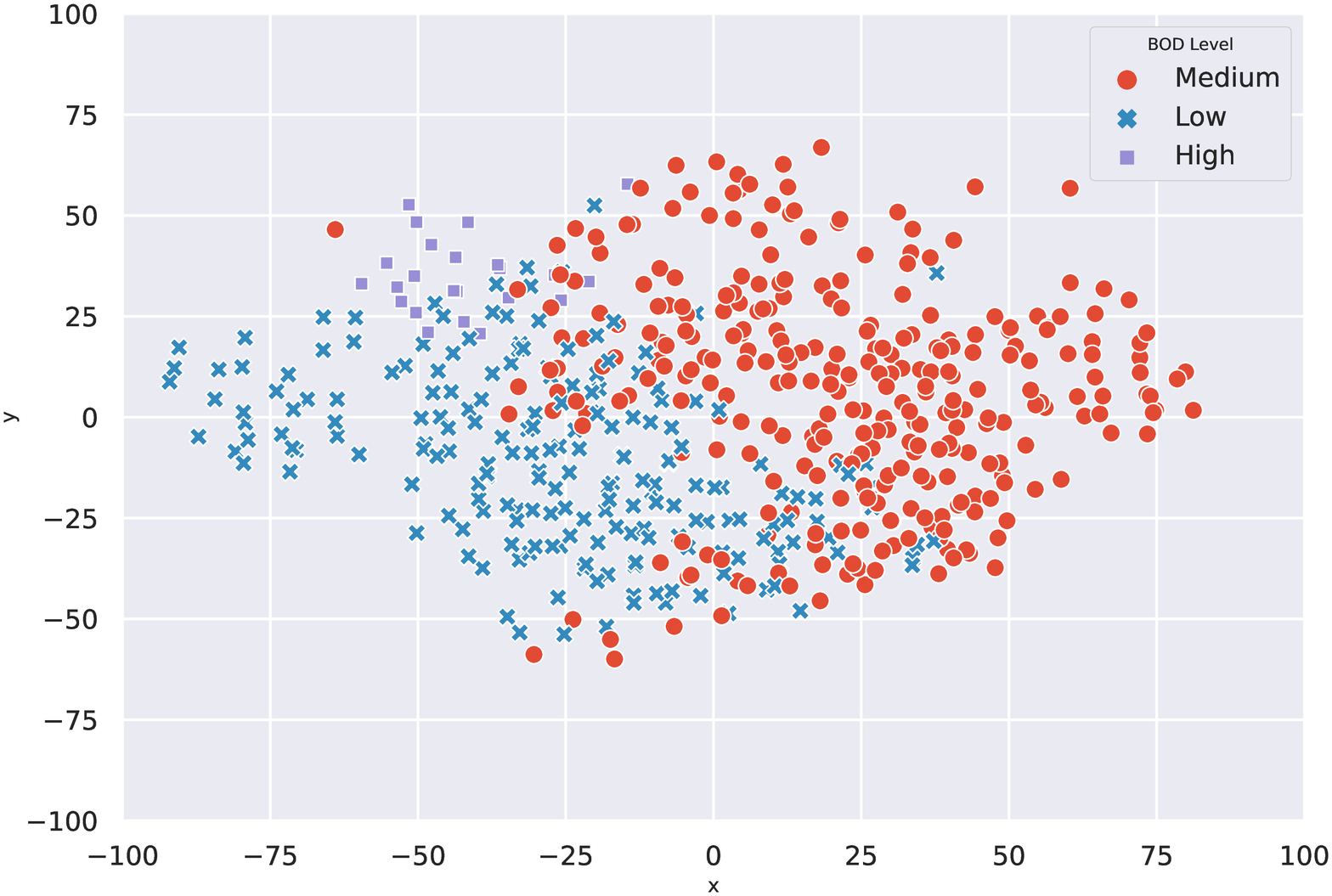}
  \caption{t-SNE with the target being BOD Levels}
  \label{fig:sub8}
\end{subfigure}
\caption{t-Distributed Stochastic Neighbour Embedding. From (a) it is clear that the data fall into three major clusters based on the BOD values: 0-0.2,0.4-0.6 and 0.8-1.0, which can be clustered as: Low, Medium and High BOD Levels, hence we cluster and categorize  as in (b). }
\label{fig:test3}
\end{figure*}

Table 2, summarizes the  results of Pearson correlation analysis among all the variables in the dataset. Table 3 further summarizes the results in relation to the variable of interest. From the results in Table 3, it is clear that three of the parameters in association with BOD had at least medium to strong level of association. These are:  Dissolved Oxygen (-0.522), Nitrogen (0.285) and Fecal Coliform (0.299). In addition to this, Total coliforms showed small but significant strength of correlation with BOD (0.174). Temperature, Conductivity and pH all showed very low strength of association with BOD $(\le|0.1|)$. These findings agreed with the results of PCA analysis above and informed the discarding of temperature, Conductivity and pH in the training of the model. 

The findings by Abyaneh, showed a high significance of pH in prediction of BOD\cite{Abyaneh2014}. However, these results showed that BOD is least sensitive to pH among other parameters in this study. This difference in the two findings can be attributed to the difference in the parameters under review in this study and those selected by Abyaneh. Abyaneh chose: Total Suspended Solids (TSS), Temperature(T), Total Suspended (TS) and pH for the study. The present  study on the other hand has considered DO, T, pH, Conductivity, Nitrogen, Fecal Coliforms and Total Coliforms.  As noted however by Abyaneh, the type of input parameters is key in this process and therefore, the difference between these findings can be justified by the difference in the parameters chosen for the two studies. Dissolved Oxygen (DO) was found to have a strong negative correlation (-0.522) which is indicative of high inverse correlation between Dissolved Oxygen and BOD. Dogan et al\cite{Dogan2009} notes that the effects of excess BOD results in low dissolved oxygen concentration in water and hence unsuitable life conditions for flora and fauna in the water. This finding confirms the results of this study.

\begin{figure*}[htb!]
\centering
\begin{subfigure}{0.5\textwidth}
  \centering
  \includegraphics[width=\textwidth]{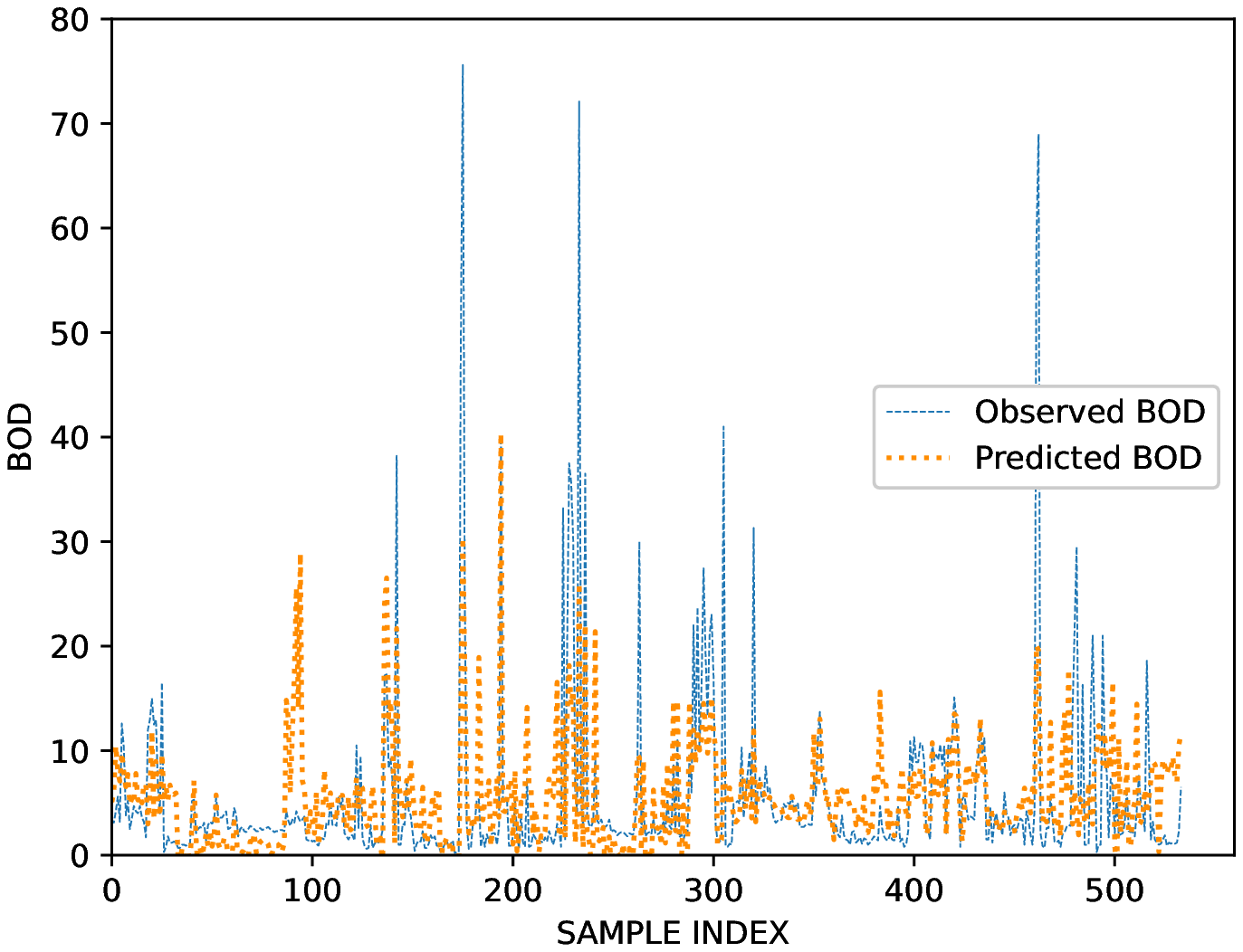}
  \caption{}
  \label{fig:sub3}
\end{subfigure}%
\begin{subfigure}{0.5\textwidth}
  \centering
  \includegraphics[width=\textwidth]{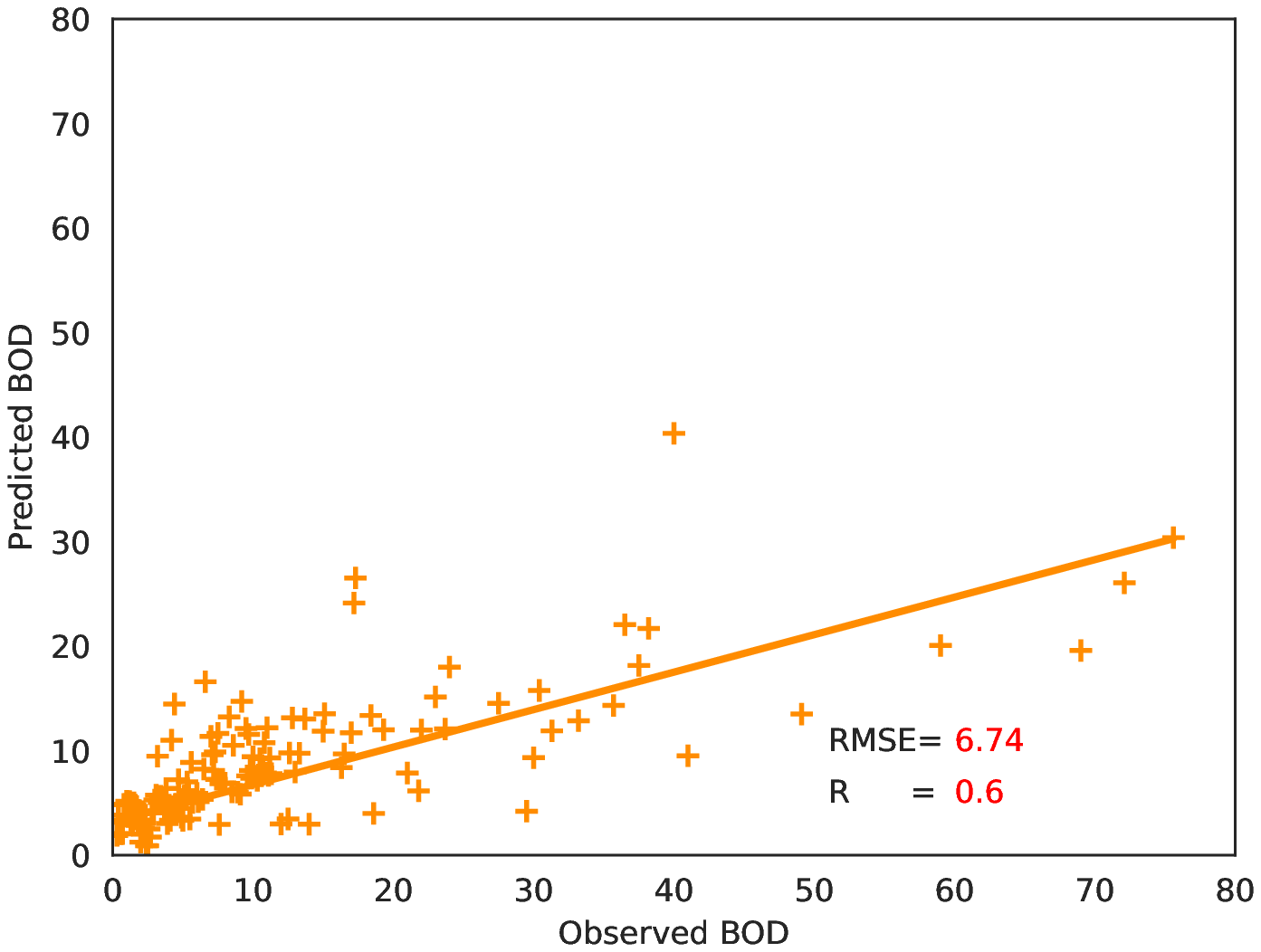}
  \caption{}
  \label{fig:sub4}
\end{subfigure}
\caption{Comparison of observed and predicted values. (a) Hydrograph of the predicted BOD versus the observed BOD and (b) Scatter plot of the predicted BOD versus the observed BOD. The model performance was evaluated with  90\%/10\% training/test split ratio and  87.5\% accuracy was achieved.}
\label{fig:test1}
\end{figure*}

\begin{figure*}[t!]
\centering
\begin{subfigure}{0.5\textwidth}
  \centering
  \includegraphics[width=\textwidth]{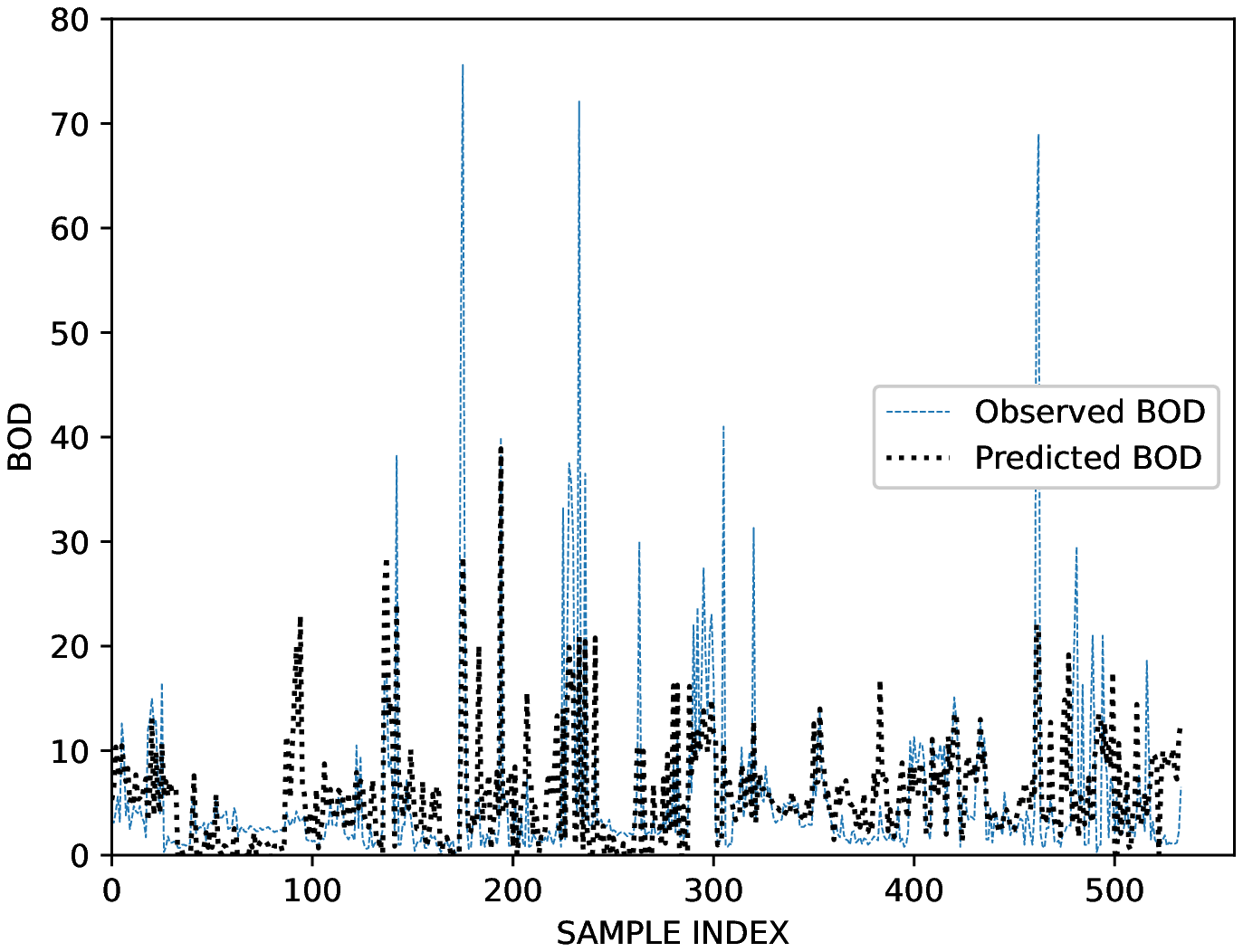}
  \caption{}
  \label{fig:sub5}
\end{subfigure}%
\begin{subfigure}{0.5\textwidth}
  \centering
  \includegraphics[width=\textwidth]{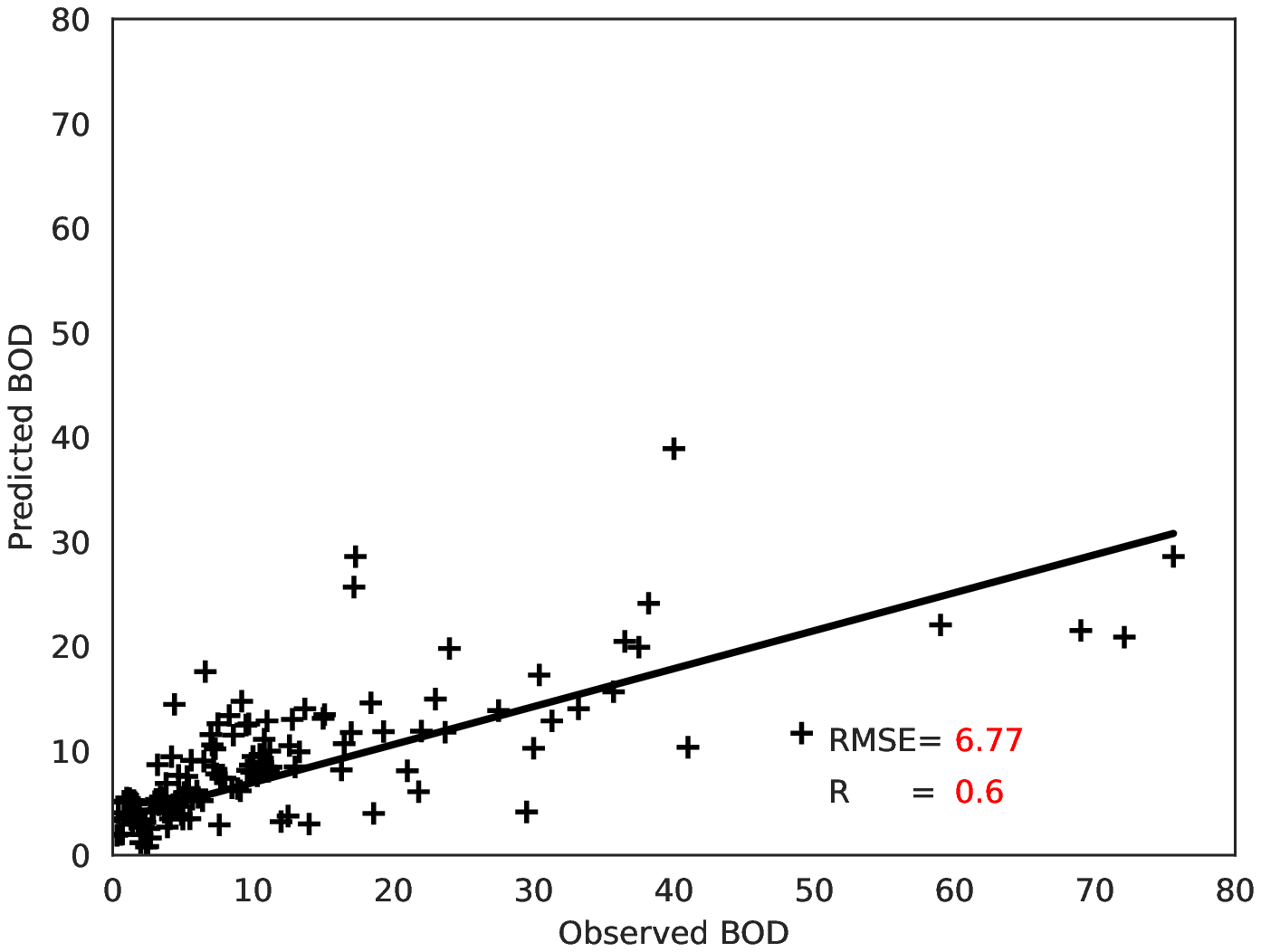}
  \caption{}
  \label{fig:sub6}
\end{subfigure}
\caption{Comparison of observed and predicted values. (a) Hydrograph of the predicted BOD versus the observed BOD and (b) Scatter plot of the predicted BOD versus the observed BOD. The model performance was evaluated with  80\%/20\% training/test split ratio and 70.3\% accuracy was achieved.}
\label{fig:test2}
\end{figure*}

The accuracy of the two models obtained from the data split of 90\%/10\% and 80\%/20\%, when tested were both excellent as they both fell within the acceptable industrial standard of 70\%-90\%\cite{Barkved2022}. The 90\%/10\% split of data for the training/test set achieved the highest accuracy of 87.5\%. On the other hand, the 80\%/20\% split achieved a relatively lower percentage accuracy of 70.3\%. It is evident from this that the accuracy of a model is increased by increasing the ratio of the training/test set of the data.

A hydrograph and scatter plot for the 90\%/10\% training/test data split alongside its scatter plot is shown in Figure 6. From the hydrograph (Figure 6(a)), it is clear that the estimated values of BOD closely follow the actual values. The Scatter plot (Figure 6(b)), agrees with the hydrograph. The r value is 0.60 which is an indicator of how well the model and observed values are in agreement. The RMSE value in this case is 6.74 mg/L. Similarly, the hydrograph for 80\%/20\% training/test data split (Figure 7(a)), shows the predicted BOD to closely follow the actual BOD. Similarly,the scatter plot (Figure 7(b)), confirms this. The r value from the 80\%/20\% split ratio was comparable to that of the 90\%/10\% split ratio. This indicated that a further increase in split ratio beyond 80\%/20\% does not improve the fit of the model on the data. In addition the RMSE for the 80\%/20\% data split was 6.77 mg/L, which is a negligible rise in comparison with that obtained for the 90\%/10\% split. This is an indicator of the fact that further increase in the training/test split ratio beyond 80\%/20\% does not significantly improve the prediction capacity of the MLR model. Rácz et al,\cite{Racz2021} used four split ratios of: 50\%,60\%, 70\% and 80\%. The 80\%/20\% split ratio achieved the best performance. In addition, the author  concludes that 80\%/20\% split ratio will provide enough training samples. The results of this study agree with these findings by Rácz et al. In modelling a waste water treatment plant using Artificial Neural Network (ANN) for prediction of water quality parameters, Güçlü and Dursun \cite{Guclu2010}, used 240/290 of the dataset for training and the rest as test set. This was approximately 80\%/20\% split ratio. Their  main aim in using this split ratio was to help in avoiding overfitting.

Equation (2) gives the approximated model for the 87.5\% accuracy on 90\%/10\% data split ratio, while equation (3) gives the approximated model for the 70.3\% accuracy of the model trained on 80\%/20\% split ratio. Both of these equations have the coefficients of the independent variables rounded off to 3 decimal places. The variables $x_1$, $x_2$, $x_3$ and $x_4$  are the input variables. They represent: Dissolved Oxygen (DO), Nitrogen (Nitrate\_N\_Nitrite\_N), Fecal Coliform and Total Coliform respectively.
\begin{equation}
    {y}={-2.305}_{X_1}+4.882\times10^{-1}x_2+
    8.939\times10^{-5}x_3  
    +2.468\times10^{-7}x_4+ 18.792
\end{equation}

\begin{equation}
    {y}={-2.187}_{X_1}+5.292\times10^{-1}x_2+
    1.094\times10^{-4}x_3-1.777\times10^{-7}x_4+ 17.781
\end{equation}

Abyaneh,\cite{Abyaneh2014}, predicted BOD using both MLR and ANN. For MLR model the authors found r value of 0.53 and RMSE of 37.8 mg/L and r value of 0.83 with RMSE of 25.1 mg/L when using ANN. The results of the present study showed improved ability of prediction with RMSE of 6.77 mg/L and 6.74 mg/L for the 80\%/20\% and 90\%/10\% split ratios respectively and r value of 0.60. The improved RMSE, which is an indicator of better prediction capacity is attributed to the better choice of input parameters. It is therefore clear that the performance of MLR model is highly dependent on the input parameters and better performance can be achieved with better choices of the input parameters.  A number of research work have shown that Artificial Neural Network (ANN) based models produces better performance than linear regression models in water quality prediction\cite{Heddam2016,Ouma2020,Hamada2018,Zhu2018}. However, some work too have indicated that Linear Regression models perform better in prediction of water quality parameters than Artificial Neural Network models\cite{Sivakumar2008}. This work when compared with some of the results of the work done previously, has shown that MLR model can a times perform better than ANN model depending on the choice of input parameters. However, absolute conclusion on whether MLR can perform better than ANN in this case scenario will require development of ANN model using the same input parameters of DO, Nitrogen, Fecal Coliform and Total Coliform to predict BOD. This however is beyond the scope of this work.

\section{Conclusion}
Multivariate Linear Regression Model was used to construct linear model for BOD prediction on the basis of four water quality parameters namely: Dissolved Oxygen (DO), Nitrogen (Nitrate\_N\_Nitrite\_N), Fecal Coliforms and Total Coliforms with two sets of training/test split ratios of the data. In both cases, the model had high statistical quality with low prediction error (RMSE=6.74 mg/L and 6.77 mg/L for 90\%/10\% and 80\%/20\%  training/test data split ratios respectively), with a good fit (r=0.60). The accuracy on testing too was excellent. The 90\%/10\% training/test data split ratio achieved 87.5\% accuracy and the 80\%/20\% training/test split ratio achieved 70.3\% accuracy. Both of these accuracies were within the acceptable industrial standards. 

The main conclusions of this study are threefold:
\begin{enumerate}[(i)]

\item Among the key water quality parameters, there is  a strong correlation between Dissolved Oxygen, Fecal Coliforms, Total Coliforms and Nitrogen to BOD.

\item	80\%/20\% training/test data split ratio is the optimum ratio in the training of a model and although a high ratio above this may improve the accuracy of the model in regards to the test data, the improvement on the prediction capacity or the fitness of the model to the data is negligible. 

\item	Depending on the parameters chosen, MLR model can perform better than ANN models in scenarios where the chosen parameters used in MLR have better correlation to the target variable than those parameters chosen for use in ANN models. 

\end{enumerate}

We note that MLR is a good and straightforward technique for prediction of BOD in waste water treatment plants. Its application will greatly improve the performance of the Waste Water Treatment Plants by providing shorter time of BOD prediction. This will aid in decision making in regards to the appropriate treatment process of the waste water. 

In our future work, we aim to apply Neural Networks analysis on the same dataset and the same input variables, to help in providing the comparison  of the neural network and the MLR models as presented in this work.

\bibliographystyle{ACM-Reference-Format}
\bibliography{sample-base}

\end{document}